\documentstyle[12pt]{article}

\makeatletter                                                           %LMS

\def\section{\@startsection {section}{1}{\z@}{-1.5ex plus -.5ex         %LMS
minus -.2ex}{1ex plus .2ex}{\large\bf}}                                 %LMS

\def\@thmcountersep{}                                                   %LMS

\long\def\@makecaption#1#2{\vskip 10pt \setbox\@tempboxa\hbox{#1. #2}   %LMS
   \ifdim \wd\@tempboxa >\hsize   % IF longer than one line:            %LMS
       #1. #2\par                 %   THEN set as ordinary paragraph.   %LMS
     \else                        %   ELSE  center.                     %LMS
       \hbox to\hsize{\hfil\box\@tempboxa\hfil}                         %LMS
   \fi}                                                                 %LMS

\def\ps@headings{                                                       %LMS
 \def\@oddhead{\footnotesize\rm\hfill\runninghead\hfill}                %LMS
 \def\@evenhead{\@oddhead}                                              %LMS
 \def\@oddfoot{\rm\hfill\thepage\hfill}\def\@evenfoot{\@oddfoot} }      %LMS

\makeatother                                                            %LMS

\title{Minisuperspaces: Symmetries and Quantization}{}{}

\def\runninghead{ASHTEKAR ET AL. :\quad SYMMETRIES AND QUANTIZATION}

\author{
{\em Abhay Ashtekar} \thanks{Physics Department, Syracuse University, Syracuse,
NY 13244-1130, USA and Inter-University Center for Astronomy and Astrophysics,
Pune, 411017, India. This author was supported in part by NSF grant
PHY90-16733, by a travel grant from the United Nations Development Office and
by research funds provided by Syracuse University.}
\and
{\em Ranjeet Tate} \thanks{Physics Department, University of
California, Santa Barbara, CA 93106-9530, USA.  This author was
supported in part by NSF grant PHY90-16733 and by research funds
provided by Syracuse University.}
\and
{\em Claes Uggla} \thanks{Physics Department, Syracuse University, Syracuse, NY
13244-1130, USA and Department of Physics, University of Stockholm,
Vanadisv\"agen 9, S-113 46 Stockholm, Sweden. This author was supported by a
grant from the Swedish National Science Research Council and by research funds
provided by Syracuse University.}
}

\date{} % no date wanted.                                              %LMS

\begin{document}

\pagestyle{headings}                                                   %LMS
\flushbottom                                                           %LMS

\maketitle
\vspace{-10pt} % include according to taste.

{\centerline{SU-GP-92/2-5, gr-qc/9302026}}
\begin{abstract}
In several of the class A Bianchi models, minisuperspaces admit symmetries. It
is pointed out that they can be used effectively to complete the Dirac
quantization program. The resulting quantum theory provides a useful platform
to investigate a number of conceptual and technical problems of quantum
gravity.
\end{abstract}

\def\ie{{i.e.}}
\def\eg{{e.g.}}
\def\three{\,^3 \kern-1.5pt}

\def\newblock{}
\newcommand{\half}{\frac{1}{2}}

% From the TeXbook, a poor man's boldface in math mode:
\def\pmb#1{\setbox0=\hbox{$#1$}%
  \kern-.025em\copy0\kern-\wd0
  \kern.05em\copy0\kern-\wd0
  \kern-.025em\raise.0433em\box0}
% and reducing by a factor of .7 for \scriptstyle:
\def\pmbs#1{\setbox0=\hbox{$\scriptstyle #1$}%
  \kern-.0175em\copy0\kern-\wd0
  \kern.035em\copy0\kern-\wd0
  \kern-.0175em\raise.0303em\box0}

\def\bfbeta{{\pmb{\beta}}}
\def\bfsbeta{\pmbs{\beta}}

\def\fraction#1#2{{\textstyle{#1\over#2}}} \def\fr{\fraction}
\def\sfraction#1#2{{\scriptstyle{#1\over#2}}} \def\sfr{\sfraction}
\def\d{\partial}

\def\vol#1{{\bf #1}}
\def\journalfont{\it}         % this allows redefinition of the font later
\def\jou#1{{\journalfont #1\ }}
\def\aaa{\jou{  Astron.\ Astrophys.}}
\def\aip{\jou{  Adv.\ Phys.}}
\def\am{\jou{   Ann.\ Math.)}}
\def\ap{\jou{   Ann.\ Phys.\ (N.Y.)}}
\def\apj{\jou{  Astrophys.\ J.}}
\def\cjp{\jou{  Can. J. Phys.}}
\def\cmp{\jou{  Commun.\ Math.\ Phys.}}
\def\cqg{\jou{  Class.\ Quantum Grav.}}
\def\grg{\jou{  Gen.\ Relativ.\ Grav.}}
\def\jmp{\jou{  J.\ Math.\ Phys.}}
\def\jpamg{\jou{ J.\ Phys.\ A: Math.\ Gen.}}
\def\mnras{\jou{ Mon.\ Not.\ R.\ Ast.\ Soc.}}
\def\nat{\jou{  Nature}}
\def\ncim{\jou{ Nuovo Cim.}}
\def\nucp{\jou{ Nuc.\ Phys.}}
\def\ncb{\jou{  Il Nuovo Cimento ``B}}
\def\pl{\jou{   Phys.\ Lett.}}
\def\pr{\jou{   Phys.\ Rev.}}
\def\prep{\jou{ Phys.\ Rep.}}
\def\prl{\jou{  Phys.\ Rev.\ Lett.}}
\def\ptp{\jou{  Prog.\ Theor.\ Phys.}}
\def\rmp{\jou{  Rev. Mod. Phys.}}
\def\spj{\jou{  Sov.\ Phys.\ JETP}}
\def\gr{04.20.-q}            % General Relativity
\def\cosmology{98.80.Dr}     % Theoretical cosmology

\def\Lie{{\cal L}}
\def\Mscr{{\cal M}}
\def\halfcone{\hbox{$\Mscr^-_{1/2}$}}
\def\cone{\hbox{$\Mscr^-$}}
\def\IP#1#2{\langle\, #1\, |\, #2\, \rangle}  %\IP\psi\phi
\def\Div#1#2{{\rm Div}_{#1}#2}                %\Div\epsilon{V}
\def\cV{{\cal V}}
\def\ps{phase space}
\def\qtzn{quantization}
\def\dd{{\rm d}}
\def\Real{{\rm I}\! {\rm R}}
%\\newcommand{\support}[1]{\thanks{#1}\ } %??

\newcommand{\bu}{\bar u}
\newcommand{\bz}{\beta^0}
\newcommand{\bp}{\beta^+}
\newcommand{\bm}{\beta^-}
\newcommand{\bbz}{\bar\beta^0}
\newcommand{\bbp}{\bar\beta^+}
\newcommand{\bbm}{\bar\beta^-}
\newcommand{\tphi}{\tilde\phi}
\newcommand{\tbz}{\tilde\beta^0}
\newcommand{\tbp}{\tilde\beta^+}
\newcommand{\tbm}{\tilde\beta^-}
\newcommand{\tp}{\tilde{p}}
\newcommand{\tpz}{{\tilde p}_0}
\newcommand{\tpp}{{\tilde p}_+}
\newcommand{\tpm}{{\tilde p}_-}
\newcommand{\bpz}{{\bar p}_0}
\newcommand{\bpp}{\bar{p}_+}
\newcommand{\bpm}{\bar{p}_-}

\newcommand{\be}{\begin{equation}}
\newcommand{\ee}{\end{equation}}

%%%%%%%%%%%%%%%%%%%%%%%%%%%%%%%%%%%%%%%%%%%%%%%%%%%%%%%%%%%%%%%%%%%%%%%%%%%%%%%
%%%%%%%%%%%%%%%%%%%%%%%%%%%%%%%%%%%%%%%%%%%%%%%%%%%%%%%%%%%%%%%%%%%%%%%%%%%%%%%

% To use \eqalign in LaTeX

% We make @ signs act like letters, temporarily, to avoid conflict
% between user names and internal control sequences of plain format.
\catcode`@=11

\newdimen\jot \jot=3pt
\newskip\z@skip \z@skip=0pt plus0pt minus0pt
\newdimen\z@ \z@=0pt % can be used both for 0pt and 0
\dimendef\dimen@=0

\def\m@th{\mathsurround=\z@}

\def\ialign{\everycr{}\tabskip\z@skip\halign} % initialized \halign

\def\openup{\afterassignment\@penup\dimen@=}
\def\@penup{\advance\lineskip\dimen@
  \advance\baselineskip\dimen@
  \advance\lineskiplimit\dimen@}

\def\eqalign#1{{
% longer lineskips in displays (TeX book p.78)
%\baselineskip=20pt \lineskip=10pt \lineskiplimit=4pt
\null\,\vcenter{\openup\jot\m@th
  \ialign{\strut\hfil$\displaystyle{##}$&$\displaystyle{{}##}$\hfil
      \crcr#1\crcr}}\,} }
%-------------------------------------------------------------------------
% while we're at it why not a multiple point alignment:
\def\meqalign#1{\null\,\vcenter{\openup\jot\m@th
  \ialign{\strut\hfil$\displaystyle{##}$&&$\displaystyle{{}##}$\hfil
      \crcr#1\crcr}}\,}
%-------------------------------------------------------------------------
\catcode`@=12   % back to nonletter category

%%%%%%%%%%%%%%%%%%%%%%%%%%%%%%%%%%%%%%%%%%%%%%%%%%%%%%%%%%%%%%%%%%%%%%%%%%%%%%%
%%%%%%%%%%%%%%%%%%%%%%%%%%%%%%%%%%%%%%%%%%%%%%%%%%%%%%%%%%%%%%%%%%%%%%%%%%%%%%%

\section{Introduction}

Minisuperspaces are useful toy models for canonical quantum gravity because
they capture many of the essential features of general relativity and are at
the same time free of the technical difficulties associated with the presence
of an infinite number of degrees of freedom. This fact was recognized by
Charlie Misner quite early and, under his leadership, a number of insightful
contributions were made by the Maryland group in the sixties and the seventies.
Charlie's own papers are so thorough and deep that they have become classics;
one can trace back to them so many of the significant ideas in this area.
Indeed, it is a frequent occurrence that a beginner in the field gets excited
by a new idea that seems beautiful and subtle only to find out later that
Charlie was well aware of it. It is therefore a pleasure and a privilege to
contribute an article on minisuperspaces to this Festschrift --of course,
Charlie himself may already know all our results!

In this paper we shall use the minisuperspaces associated with Bianchi models
to illustrate some techniques that can be used in the quantization of
constrained systems --including general relativity-- and to point out some of
the pitfalls involved. We will carry out canonical quantization and impose
quantum constraints to select the physical states {\it a la} Dirac. However, as
is by now well-known, the Dirac program is incomplete; it provides no
guidelines to introduce the inner product on the space of these physical
states. For this, additional input is needed. A possible strategy \cite{k:cm}
is to use appropriate symmetries of the classical theory.

Consider for example a ``free'' particle of mass $\mu$ moving in a stationary
space-time (where the Killing field is everywhere timelike). The constraint
$P^aP_a +\mu^2 =0$ implies that the physical states in the quantum theory must
satisfy the Klein-Gordon equation $\nabla^a\nabla_a \phi -\mu^2 \phi =0$. To
select the inner product  on the space of physical states, one can use the fact
\cite{am:qf,k:cmp} that the space of solutions admits a unique  K\"ahler
structure which is left invariant by the action of the timelike Killing field.
The isometry group of the underlying spacetime is unitarily implemented on the
resulting Hilbert space; the classical symmetry is promoted to a symmetry of
the quantum theory. In fact there is a precise sense \cite{aa:sen} in which
this last condition selects the Hilbert space structure of the quantum theory
uniquely. We will see that a similar strategy is available for all class A
Bianchi models except types VIII and IX. In these models, therefore, the Dirac
program can be completed and the resulting mathematical framework can be used
to test many of the ideas that have been proposed in quantum cosmology. For
example, one can investigate if the resulting Hilbert space admits a preferred
wave function which can be taken to be the ground state and hence a candidate
for the ``wave function of the universe''. We shall find that, in general, the
answer is in the negative.

The paper is organized as follows. In section 2, we recall some of the key
features of the mathematical description of Bianchi models. In section 3, we
point out that in type I, II, VI$_0$ and VII$_0$ models, the supermetric on
each minisuperspace admits a null Killing vector whose action leaves the
potential term in the scalar (or, Hamiltonian) constraint invariant. Such
Killing fields are called {\it conditional symmetries} \cite{kk:jmp}. We then
quantize these models by requiring that the symmetries be promoted to the
quantum theory. We conclude in section 4 with a discussion of the ramifications
of these results for quantum gravity in general and quantum cosmology in
particular.

\section{Hamiltonian cosmology}

In this paper, we shall consider spatially homogeneous models which admit a
three-dimensional isometry group which acts simply and transitively on the
preferred spatial slices. (Thus, we exclude, e.g., the Kantowski-Sachs models.)
In this case, one can choose, on the homogeneous slices, a basis of (group
invariant) 1-forms
$\omega^a\> (a=1-3)$ such that
 \be
    d\omega^a=-\fr12 C^a{}_{bc} \omega^b\wedge\omega^c\ ,
 \ee
where $C^a{}_{bc}$ are the components of the structure constant tensor of the
Lie algebra of the associated Bianchi group. The vanishing or nonvanishing of
the trace $C^b{}_{ab}$ divides these models into two classes called class A and
class B respectively. This classification is important for quantization because
while a satisfactory Hamiltonian formulation is available for all class A
models, the standard procedure for obtaining this formulation fails in the case
of general class B models (although the possibility that a modified procedure
may eventually work is not ruled out. See, e.g., the suggestion made in the
last section of \cite{as:bia}.) Since the Hamiltonian formulation is the point
of departure for canonical quantization, from now on we will consider only the
class A models.

Let us simplify matters further by restricting ourselves to those spatially
homogeneous metrics which can be diagonalized. (This can be achieved in the
type VIII and IX models by exploiting the gauge freedom made available by the
vector or the diffeomorphism constraint. In the remaining models there {\it is}
a loss of generality, which, however, is mild in the sense that the restriction
corresponds only to fixing certain constants of motion. For details, see
\cite{as:bia}.) Thus, we consider space-time metrics of the form:
 \be
  ^{(4)}ds^2 = -N^2(t)dt^2 + \sum_{a=1}^3 g_{aa}(t) (\omega^a)^2\ ,
 \ee
where $N(t)$ is the lapse function (see, e.g., \cite{m:ani}). Since the trace
of the structure constants vanishes, they can be expressed entirely in terms of
a symmetric, second rank matrix $n^{ab}$:
 \be
   C^a{}_{bc} = \epsilon_{mbc}n^{ma},
 \ee
where $\epsilon_{mbc}$ is a 3-form on the 3-dimensional Lie algebra. The
signature of $n^{am}$ can then be used to divide the class A models into
various types. In the literature, one generally uses the basis which
diagonalizes $n^{am}$ and then expresses $C^a{}_{bc}$ as \cite{j:uni}:
 \be
     C^a{}_{bc} = n^{(a)} \epsilon_{abc} \qquad ({\rm no\ sum\ over}\ a)\ .
 \ee
The constants $n^{(a)}$ are then used to characterize the different class A
Bianchi types. A convenient set of choices is given in the following table
\cite{j:uni}:

\vskip0.5truecm
\renewcommand{\arraystretch}{2}
\centerline{
 \begin{tabular}{||c||c|c|c|c|c|c||} \hline
 %         & \multicolumn{6}{c||}{Bianchi types}
 %                                                           \\ \hline\hline
            &I  & II&VI$_0$&VII$_0$&VIII&IX  \\\hline
  $n^{(1)}$ &0  & 1 &  1   &   1   & 1  &1 \\\hline
  $n^{(2)}$ &0  & 0 & $-$1 &   1   & 1  &1 \\\hline
  $n^{(3)}$ &\hphantom{-}0  & \hphantom{-}0 &  \hphantom{-}0  & \hphantom{-}0
   &$-$1 & \hphantom{-}1 \\\hline
 \end{tabular}
}
\vskip0.5truecm

To simplify the notation and calculations, let us use the Misner \cite{m:mini}
parametrization of the diagonal spatial metric in (2)
 \be \eqalign{
   {\bf g} & \equiv g_{ab} = {\rm diag}(g_{11}, g_{22}, g_{33}) \equiv
   e^{2{\bfsbeta}}\ .\cr
   {\bfbeta} &= \bz{\rm diag}(1, 1, 1) +
                  \bp{\rm diag}(1, 1, -2) +
                  \bm{\rm diag}(\sqrt3, -\sqrt3 , 0)\ .
 }\ee
This parametrization leads to conformally inertial coordinates for the Lorentz
(super) metric defined by the ``kinetic term'' in the scalar constraint. For
Bianchi types I, II, VI$_0$ and VII$_0$, the scalar constraint (which,
incidentally, can be obtained directly by applying the ADM procedure
\cite{adm:adm} to {\it all} class A models \cite{j:uni}) now takes the form
 \be
  C(\beta^A,p_A):= \fr{1}{24} N e^{-3\bz} \eta^{AB} p_A p_B +
      \fr12 N e^{\bz + 4\bp} \left( n^{(1)} e^{2\sqrt3\bm}
              -n^{(2)}e^{-2\sqrt3\bm} \right)^2 = 0 .
 \ee
Here the upper case latin indices $A, B ...$ range over $0,+,-$; $p_A$ are the
momenta canonically conjugate to $\beta^A$ and the matrix $\eta^{AB}$ is given
by diag$(-1, 1,1)$. To further simplify matters, one usually chooses the ``Taub
time gauge'', i.e.\ one chooses the lapse function to be $N_T=
12\exp{3\beta^0}$ \cite{t:taub51}. (This choice of gauge is also known as
Misner's supertime gauge \cite{m:mini}.) With this lapse, the scalar constraint
takes the form:
 \be\eqalign{ \label{eq:ht}
  C_T &= \fr12 \eta^{AB} p_A p_B + U_T = 0\ ,\cr
  U_T &=  6e^{4(\bz + \bp)}\left( n^{(1)}e^{2\sqrt3\bm}
              - n^{(2)}e^{-2\sqrt3\bm}\right)^2\ .\cr
 }\ee
Consequently, the dynamics of all these Bianchi models is identical to that of
a particle moving in a 3-dimensional Minkowski space  under the influence of a
potential $U_T$. Finally, because we have restricted ourselves to metrics which
are diagonal, the vector constraint is identically satisfied. Thus, the
configuration space is 3-dimensional and there is one nontrivial constraint
(\ref{eq:ht}). The system therefore has two true degrees of freedom.

\section{Quantization}

To quantize these systems we must allow the two true degrees of freedom to
undergo quantum fluctuations. Following the Dirac theory of constrained
systems, let us consider, to begin with, wave functions $\phi(\vec{\beta})$
where $\vec\beta\equiv (\beta^0, \beta^+, \beta^- )$. The physical states can
then be singled out by requiring that they should satisfy the quantum version
of the constraint equation (\ref{eq:ht}):
 \be\eqalign{ \label{eq:kg}
   & \Box \phi  - \mu^2 (\vec\beta) \phi = 0\ ,{\rm where} \cr
   & \Box = -\left(\frac{\partial}{\partial \bz}\right)^2 +
             \left(\frac{\partial}{\partial \bp}\right)^2 +
             \left(\frac{\partial}{\partial \bm}\right)^2\ ,
            {\rm and}\cr
   & \mu^2 (\vec \beta) = 12 e^{4(\bz + \bp)}
     \left(n^{(1)}e^{2\sqrt3\bm} - n^{(2)}e^{-2\sqrt3\bm}\right)^2\ .\cr
 }\ee
Thus, the physical states are solutions of a ``massive'' Klein-Gordon equation
where the mass term, however, is ``position dependent'': it is a potential in
minisuperspace. Our first task is to endow the space of these states with the
structure of a complex Hilbert space. It is here that we need to extend the
Dirac theory of quantization of constrained systems.

Consider, for a moment, type I models  where the potential vanishes. In this
case, we are left with just the free, massless Klein-Gordon field in a
3-dimensional Minkowski space. Therefore, to endow the space of solutions with
the structure of a Hilbert space, we can use the standard text-book procedure.
First, decompose the fields into positive and negative frequency parts (with
respect to a timelike Killing vector field) and restrict attention to the space
$V^+$ of positive frequency fields $\phi^{+}$. Then, introduce on $V^+$ the
inner-product:
 \be\label{eq:pfip}
   \langle \phi^+_1 ,\phi^+_2 \rangle := -2i \Omega (\overline{\phi^+_1},
   \phi^+_2),
 \ee
where ``overbar'' denotes complex conjugation and where $\Omega$ is the natural
symplectic structure on the space of solutions to the Klein-Gordon equation:
 \be \label{eq:symplectic}
  \Omega (\phi_1 , \phi_2) := \int_{\Sigma} d^2S^A\  (\phi_2\partial_A
   \phi_1 -\phi_1\partial_A\phi_2),
 \ee
$\Sigma$ being any (2-dimensional) Cauchy surface on $(M, \eta^{AB})$.
Alternatively, we can restrict ourselves to the vector space $V$ of {\it real}
solutions to the Klein-Gordon equation and introduce on this space a complex
structure $J$ as follows: $J\circ \phi = i(\phi^+ - \overline{\phi^+})$. This
$J$ is a real-linear operator on $V$ with $J^2 = -1$. It enables us to
``multiply'' real solutions $\phi\in V$ by complex numbers:
$(a+ib)\circ\phi:=a\phi+bJ\circ\phi$, which is again in $V$. Thus $(V,J)$ can
be regarded as a complex vector space. Furthermore, $J$ is compatible with the
symplectic structure in the sense that $\Omega (J\phi_1, \phi_2)$ is a
symmetric, positive definite metric on $V$. Therefore, $\langle .,.\rangle:=
\Omega(J.,.) - i\Omega(.,.)$ is a Hermitian inner product on the complex vector
space $(V,J)$ and thus $\{V, \Omega, J, \langle .,.\rangle\}$ is a K\"ahler
space \cite{am:qf,k:cmp,aa:sen}. There is a natural isomorphism between the
complex vector spaces $V^+$ and $\{V,J\}$: $\phi^+ = \textstyle{1\over 2}(\phi
- iJ\phi)$. Furthermore, this map preserves the Hermitian inner products on the
two spaces. Thus, the two descriptions are equivalent. While we introduced $J$
in terms of positive frequency fields, one can also proceed in the opposite
direction and treat $J$ as the basic object. One would then use the above
isomorphism to {\it define} the positive frequency fields. In fact, it turns
out that the description in terms of $\{V,J\}$ can be extended more directly to
the Bianchi models with non-vanishing potentials (as well as to other contexts
such as quantum field theory in curved space-times). This is the strategy we
will adopt.

Let us make a small digression to discuss the problem of finding the required
operator $J$ in a general context and then return to the Bianchi models. Let
us suppose we are given a real vector space%
\footnote{Throughout this paper, our aim is to convey only the main ideas.
 Therefore, we will not make digressions to discuss the subtle but often
 important issues from functional analysis. In particular, we will not specify
 the precise domains of various operators nor shall we discuss topologies on
 the infinite dimensional spaces with  respect to which, for example, the
 symplectic structures are to be continuous.}
$V$ equipped with a symplectic structure $\Omega$. Thus, $\Omega: V\otimes V
\mapsto \Real$ is a second rank, anti-symmetric, non-degenerate tensor over
$V$. A 1-parameter family of canonical transformations on $V$ consists of
linear mappings $U(\lambda)$ from $V$ to itself ($\lambda \in \Real$) such that
$\Omega(U(\lambda) \circ v, U(\lambda)\circ w )=\Omega (v,w)$ for all $v,w$ in
$V$. The generator $T$ of this family, $T := dU(\lambda)/d\lambda$, is
therefore a linear mapping from $V$ to itself satisfying $\Omega (T\circ v,w) =
-\Omega(v, T\circ w)$. It is generally referred to as an {\it infinitesimal
canonical transformation}. These operators can be regarded as symmetries on
$\{V, \Omega\}$. The mathematical question of interest to us is: can one endow
$V$ with the structure of a complex Hilbert space on which $U(\lambda)$ are
unitary operators?

Since $V$ is equipped with a symplectic structure $\Omega$, to ``Hilbertize''
$V$, it is natural to seek a complex structure $J$ on it which is compatible
with $\Omega$ in the sense discussed above; the resulting K\"ahler structure
can then provide the Hermitian inner-product on the complex vector space
$\{V,J\}$. The issue of existence and uniqueness of such complex structures was
discussed in detail in \cite{aa:sen} and we will only report the final result
here. The generating function $F_T(v)$ of the canonical transformation under
consideration is simply $F_T(v) := \textstyle{1\over 2} \Omega(v, Tv)$. If this
is positive definite, then there exists a unique complex structure $J$ which is
compatible with $\Omega$ such that $U(\lambda)$ are unitary operators on the
resulting K\"ahler space: $\langle U(\lambda)\circ v, U(\lambda)\circ w)\rangle
= \langle v, w\rangle$, where, as before, the inner-product is given by
 \be\label{eq:pip}
   \langle v,w\rangle = \Omega (Jv, w) - i \Omega(v,w)\ .
 \ee
Thus, if one has available a preferred canonical transformation on the real
symplectic space $\{V,\Omega \}$ whose generating function is positive
definite, one can endow it unambiguously with the structure of a complex
Hilbert space.

The preferred complex structure is defined as follows. Using $T$, let us first
introduce (only as an intermediate step) a fiducial, real inner-product $(.,.)$
on $V$ as follows:
 \be\label{eq:fip}
    (v,w) := \fr12\Omega(v,T\circ w).
 \ee
This is indeed an inner product: it is symmetric because $T$ is an
infinitesimal canonical transformation and the resulting norm is positive
definite because it equals the generating functional $F_T(v)$. Let $\bar{V}$
denote the Hilbert space obtained by Cauchy completing $V$ w.r.t. $(.,.)$. It
is easy to check that $T^2$ is a symmetric, negative operator on $\bar{V}$,
whence it admits a self-adjoint extension which we also denote by $T^2$. Then,
 \be
   J:= -(-{T}^2)^{\textstyle{-{1\over 2}}} \cdot {T}
 \ee
is a well-defined operator with $J^2 = -1$. Using the expression (\ref{eq:fip})
of the inner-product on $\bar{V}$, it is easy to check that $J$ is indeed
compatible with $\Omega$. We can therefore introduce a new {\it Hermitian}
inner product $\langle .,.\rangle$ on the {\it complex} vector space $(\bar{V},
J)$ via (\ref{eq:pip}). The Cauchy completion $H$ of this complex pre-Hilbert
space is then the required Hilbert space of physical states. (Thus, the
inner-product (\ref{eq:fip}) and the resulting real Hilbert space $\bar{V}$
were introduced only as mathematical tools to enable us to construct the
physical Hermitian structure.) Finally, it is easy to show that $J$ commutes
with $U(\lambda)$. It then follows, from the fact that $U(\lambda)$ are
canonical transformations and from the definition (\ref{eq:pip}) of the
Hermitian inner-product, that $U(\lambda)$ are unitary. The proof of uniqueness
of $J$ is straightforward but more involved \cite{am:qf,k:cmp,aa:sen}.

In the case when $V$ is the space of solutions to the Klein-Gordon equation in
Minkowski space-time (as is the case in the Bianchi type I models) the standard
complex structure (corresponding to the positive/negative frequency
decomposition) can be obtained by choosing for $T$ the operator ${\cal L}_t$
where $t^A$ is any time translation Killing field in Minkowski space. More
generally, one can think of $J$ as arising from ``a positive and negative
frequency decomposition defined by $T$''. More precisely, $J$ is the unitary
operator in the {\it polar decomposition} \cite{rs:fun} of the operator $T$ on
the Hilbert space $\bar{V}$.

With this general machinery at hand, we can now return to the Bianchi types II,
VI$_0$, VII$_0$. Denote by $V$ the space of real solutions $\phi$ to the
operator constraint equation  (\ref{eq:kg}). Elements of $V$ represent the
physical quantum states of the system. As observed earlier, $V$ is naturally
equipped with a symplectic structure $\Omega$ (see (\ref{eq:symplectic})) and
our task is to find a compatible complex structure $J$. For this, we need a
preferred canonical transformation on $V$ with a positive generating function.
The situation in type I models suggests that we attempt to construct this
transformation using an appropriate Killing field $t^A$ of the flat supermetric
$\eta^{AB}$ on the 3-dimensional mini-superspace. However, now, the constraint
equation (\ref{eq:kg}) involves a nontrivial potential term
$\mu^2(\vec{\beta})$. Hence, for ${\cal L}_t$ to be a well-defined operator on
$V$ -- i.e, for ${\cal L}_t\,\phi$ to be again a solution to (\ref{eq:kg}) --
it is essential that ${\cal L}_t\,\mu^2 = 0$. As mentioned earlier, such a
Killing field is called a {\it conditional symmetry} \cite{kk:jmp}.
Fortunately, the models under consideration do admit such a conditional
symmetry%
\footnote{The origin of this symmetry can be traced back to the fact that each
 of the type I--VII$_0$ models admits a diagonal automorphism
 \cite{j:uni,ruj:vac} and --like all vacuum models--  enjoys a scale
 invariance.}.
An inspection of the potential term $\mu^2(\vec{\beta})$ shows that it is
invariant under the action of the diffeomorphism generated by the null Killing
field
 \be\label{eq:nkvf}
  t^A:= \left(\frac{\d}{\d\bz}\right)^A - \left(\frac{\d}{\d\bp}\right)^A
 \ee
of the supermetric $\eta^{AB}$. The question therefore reduces to that of
positivity of the generating functional of the canonical transformation
$T\circ\phi:= {\cal L}_t\ \phi$. A straightforward calculation shows that the
generating functional $F_T(\phi) \equiv \textstyle{1\over 2}\Omega (\phi,
T\phi)$ is simply the ``conserved energy'' associated with the null Killing
field:
 \be\label{eq:gen}
  F_T(\phi) = \fr12 \int_\Sigma [\partial_A \phi \partial_B \phi -
              \eta_{AB}(\partial^C\phi \partial_C\phi  + \mu^2
              \phi^2)] t^A dS^B\ ,
 \ee
where $\Sigma$ is a spacelike Cauchy surface. Since the potential is
nonnegative and $t^A$ is everywhere future-directed, it follows that
$F_T(\phi)$ is indeed positive. Hence, by carrying out a polar decomposition of
the operator $T \equiv {\cal L}_t$, we obtain a K\"ahler structure on $V$ which
provides us with the required complex Hilbert space of physical states of the
system. On this Hilbert space, the 1-parameter family of diffeomorphisms
generated by the conditional symmetry $t^A$ is unitarily implemented.

To summarize, because Bianchi types I, II, VI$_0$ and VII$_0$ admit conditional
symmetries whose generating functions are positive definite, in these cases,
one {\it can} satisfactorily complete the Dirac quantization program. In the
more familiar language of positive and negative frequency decomposition, the
Hilbert space of physical states consists of positive frequency solutions
$\phi^+$ of the quantum constraint equation (\ref{eq:kg}): $\phi^+ =
\textstyle{1\over 2}(\phi - iJ\circ\phi)$ where $\phi$ is a real solution and
$J = -(-T^2)^{-\textstyle{1\over 2}}\cdot T$. The inner product (\ref{eq:pfip})
is constructed from the familiar probability current, i.e., the symplectic
structure (\ref{eq:symplectic}): $\langle\phi^+_1, \phi^+_2\rangle= -2i\Omega
(\overline{\phi^+_1}, \phi^+_2)$. Finally, the 1-parameter family of
diffeomorphisms generated by $t^A$ is unitarily implemented in this Hilbert
space. Motions along the integral curves of $t^A$ can be interpreted  as ``time
evolution'' in the classical theory {\it both} in the minisuperspace {\it and}
in space-times: $t^A$ is future directed in the minisuperspace and, in any
space-time defined by the classical field equations, $\beta^0$ --an affine
parameter of $t^A$-- increases monotonically with physical time. In the quantum
theory, this time evolution is unitary.

This mathematical structure can now be used to probe various issues in quantum
cosmology. We conclude with an example.

Let ask whether there is a ``preferred'' state in the Hilbert space which can
be taken to be the ground state. The question is well-posed because on the
physical Hilbert space there is a well-defined Hamiltonian which generates
time-evolution in the sense described above. On real solutions to the quantum
constraint, the Hamiltonian is given by $\hat{H}\circ\phi=J\hbar\Lie_t\phi$;
while in the positive frequency representation of physical states it is given
by $\hat{H}\circ\phi^+=i\hbar\Lie_t\phi^+$. We will show that $\hat{H}$ is a
non-negative operator with zero only in the {\it continuous} part of its
spectrum. This will establish that $\hat{H}$ does not admit a (normalizable)
ground state. Thus, the most direct procedure to select a ``preferred'' state
fails in all these models. Furthermore, a detailed examination of the
mathematical structures involved suggests that as long as the configuration
space -- spanned by $\beta^\pm$ -- is noncompact, there is in fact {\it no}
preferred state in the physical Hilbert space. It would appear that to obtain
such a state, which could be taken, e.g., as the wavefunction of the universe
in these models, one must modify in an essential way the broad quantization
program proposed by Dirac.

Finally, we outline the proof of the technical assertion made above. Note
first, that the expectation value of $\hat{H}$ in any physical state is given
by
 \be\label{eq:energy}
  {\langle\phi,\hat{H}\phi\rangle\over\langle\phi,\phi\rangle} = \hbar
  {\Omega(J\phi,J\Lie_t\phi)\over\Omega(J\phi,\phi)} = \hbar {(\phi,\phi)\over
  (\phi,(-T^2)^{-\half}\phi)}\ge0,
 \ee
where we have used the expression (\ref{eq:pip}) of the physical, Hermitian
inner product $\langle.,.\rangle$, the expression (\ref{eq:fip}) of the
fiducial inner product $(.,.)$ and the fact that $(-T^2)$ is a positive
definite operator with respect to $(.,.)$. Since the expectation values of
$\hat{H}$ are non-negative, it follows that its spectrum is also non-negative.
To show that zero is in the continuous part of the spectrum, it is convenient
to work with the real Hilbert space $\bar{V}$ defined by the inner product
$(.,.)$. Using the fact that the potential $\mu^2(\vec{\beta})$ in
(\ref{eq:kg}) is smooth, non-negative and takes values that are arbitrarily
close to zero, one can show that the expectation values of $(-T^2)$ on
$\bar{V}$ can also approach arbitrarily close to zero. Hence zero is in the
spectrum of $(-T^2)$, which implies that the spectrum of $(-T^2)^{-\half}$ is
unbounded above. It therefore follows from (\ref{eq:energy}) that one can find
physical states $\phi$ in which the expectation values of $\hat{H}$ are
arbitrarily close to zero. Hence zero is in the spectrum of $\hat{H}$. Finally,
if it were in the discrete part of the spectrum, there would exist a physical
state $\phi_0$ which is annihilated by $\Lie_t$. In this state, in particular,
$\langle\phi_0,\hat{H}\phi_0\rangle = 2\hbar(\phi_0,\phi_0)$ must vanish.
However, it is straightforward to show that
 \be
  2(\phi_0,\phi_0)=\int_\Sigma\>\left( (\Lie_t\phi_0)^2+\left(\frac{\d\phi_0}
  {\d\beta^-}\right)^2 + \mu^2(\vec{\beta})\phi_0^2\right)\>d^2x
 \ee
where $\Sigma$ is any $\beta^0=const.$ 2-plane. Hence $(\phi_0,\phi_0)$ can
vanish iff $\phi_0$ and $\Lie_t\phi_0$ vanish on $\Sigma$, i.e., iff $\phi_0$
is the zero solution to (\ref{eq:kg}). Hence zero must belong only to the
continuous part of the spectrum of $\hat{H}$.

\section{Discussion}

In the literature on spatially homogeneous quantum cosmology, the emphasis has
been on type I and type IX models. Type I is the simplest and its quantization
has been well-understood for sometime now. In fact, the Hamiltonian description
of this model is the same as that of the strong coupling limit of general
relativity ($G_{\rm Newton} \mapsto \infty$), whence its quantization has been
discussed also in the context of this limit. Type IX is the most complicated of
the spatially compact models and exhibits, in a certain well-defined sense,
chaotic behavior  in the classical theory%
\footnote{The type VIII model has many of the interesting features of the
 type IX. However, it seems not to have drawn as much attention in quantum
 cosmology because it is spatially open. Many of the remarks that follow
 on the type IX model are applicable to the type VIII model as well.}.
It is also the most ``realistic'' of the Bianchi models as far as dynamics of
general relativity in strong field regions near singularities is concerned. In
the early work on quantum cosmology therefore, the focus of attention was type
IX models; type I was studied mainly as a preliminary step towards the quantum
theory of type IX. Unfortunately, the ``potential'' in the scalar constraint of
type IX models is so complicated that quantization attempts met with only a
limited success. In particular, until recently, not a single exact solution to
the quantum constraints was known, whence one could not even begin to address
the issues we have discussed in this paper%
\footnote{Over the past two years, a handful of solutions have been
 obtained \cite{kod,mr} using new canonical variables \cite{a:new} in terms
 of which the potential term disappears. While this development does
 represent progress, the solutions obtained thus far are rather special
 and there are too few to construct a useful Hilbert space.}.

The reason we could make further progress in this paper is that we analysed the
{\it intermediate models} in some detail. Although they are not as
``realistic'' as type IX, we have seen that these models do have an interesting
structure. In particular, unlike in the type I models, the potential in the
scalar constraint is non-zero, whence their dynamics is quite non-trivial
already in the classical theory. For example, we still do not know the explicit
form of a complete set of constants of motion in type VI$_0$ and type VII$_0$
space-times. In spite of this, we could complete the Dirac quantization program
because these models admit an appropriate (i.e., future-directed) conditional
symmetry. In physical terms, quantization is possible because one can define an
internal time variable on these minisuperspaces in a consistent fashion.

The resulting mathematical framework is well-suited for addressing a number of
conceptual issues in quantum gravity in general and quantum cosmology in
particular.

First is the issue of time. Using the conditional symmetry $t^A$, we were able
to construct a complex structure $J$ such that the 1-parameter family of
transformations on the physical quantum states $\phi$ induced by $t^A$ are
implemented by a 1-parameter family of unitary operators on the Hilbert space.
Hence, using the affine parameter along the vector field $t^A$ as a ``time''
variable, one can deparametrize the theory and express time evolution through a
Schr\"odinger equation: it is the introduction of the complex structure -- or,
the use of only positive frequency fields -- that provides a ``square-root'' of
the quantum scalar constraint which can be re-interpreted as the Schr\"odinger
equation. In the type I model, for example, the ``square-root'' takes the
following form:
 \be
  i\hbar\Lie_t\phi^+(\vec{\beta})=\hbar\left( (-\Delta)^\half-
  i\frac{\partial}{\partial\beta^+}\right) \phi^+(\vec{\beta})
 \ee
where $\Delta=(\partial/\partial\beta^+)^2+(\partial/\partial\beta^-)^2$ is the
Laplacian in the $\beta^\pm$ plane. Thus the argument $\beta^0$ of $\phi^+$ can
be regarded as time and $\hat{H}\equiv \hbar((-\Delta)^\half - i(\partial/
\partial\beta^+))$ can be regarded as the Hamiltonian. (The second term in the
expression for $\hat{H}$ arises simply because our evolution is along a {\it
null} Killing field (\ref{eq:nkvf}) rather than a timelike one.) Now, there
exist in the literature several distinct approaches to the issue of time in
full quantum gravity, including those that suggest that one should generalize
quantum mechanics in a way in which there is in fact {\it no} preferred time
variable. It would be fruitful indeed to apply these ideas to the
minisuperspaces considered in this paper and compare the resulting quantum
descriptions with the one obtained here using the conditional symmetry.

A second issue is the strategy to select inner products. Since we do not have
access to any symmetry group in general relativity, it has been suggested
\cite{a:book,t:diss} that we should let the ``reality conditions'' determine
the inner product. More precisely, the strategy is the following: Find a
sufficient number of real classical observables {\sl a la} Dirac --i.e.
functions on the phase space which weakly commute with the constraints-- and
demand that the inner product between quantum states be so chosen that the
corresponding quantum operators are Hermitian. In many examples, this condition
suffices to pick out the inner product uniquely. In the minisuperspaces
considered in this paper, on the other hand, we have selected the inner-product
using a completely different strategy: we exploited the fact that these models
admit a conditional symmetry of an appropriate type, thereby side-stepping the
issue of isolating a {\it complete} set of Dirac observables. It would be
interesting to identify the Dirac observables, use the ``reality conditions''
strategy and compare the resulting inner product with the one we have
introduced. (Equivalently, the question is whether the real Dirac observables
can be promoted to operators which are Hermitian with respect to the inner
product we have introduced.) This program has been completed in type I and II
models \cite{atu:tII}. While the two inner products do agree in a certain
sense, subtleties arise already in type II models. However, the question
remains open in types VI$_0$ and VII$_0$.

A third issue is the fate of classical singularities in the quantum
descriptions. Every (non-flat) classical solution belonging to the models
considered here has a singularity and one sometimes appeals to ``the rule of
unanimity'' \cite{jw:amsci} to argue that in such cases, singularities must
persist also in quantum theory. On the other hand, we have found that the
quantum evolution is given by a 1-parameter family of unitary operators%
\footnote{The situation is similar in full 2+1 gravity, where, {\it every}
 classical cosmological solution begins with a ``little bang'' where
 (there is no curvature singularity --hence the pre-fix ``little''-- but
 where) the spatial volume goes to zero. Inspite of this, the mathematical
 framework of quantum theory is complete and well-defined.}.
In all cases considered in this paper, the quantum Hilbert spaces are
well-defined and quantum evolution is unitarily implemented. This appears, at
least at first sight, to be a violation of the rule of unanimity and it is
important to understand the situation in detail. Is it the case that the
classical singularities simply disappear in the quantum theory? Or, do they
persist but in a ``tamer'' fashion? In the space-time picture provided by
classical general relativity, evolution is implemented by hyperbolic equations
which simply breakdown at curvature singularities. Is this loss of
predictability recovered in the quantum theory? That is, in quantum theory, can
we simply ``evolve through the singularity''? It would be useful to apply the
semi-classical methods available in the literature to these models both to gain
physical insight into this issue and to probe the limitations of the
semi-classical methods themselves. (Some results pertaining to these issues are
discussed in \cite{t:diss,atu:tII}.)

Finally, further work is needed to achieve a more complete understanding of the
quantum physics of these models. What we have constructed is the Hilbert space
of physical states. Any self-adjoint operator on this space may be regarded as
a quantum Dirac observable. However, unless the operator has a well-defined
classical analog it is in general not possible to interpret it physically%
\footnote{In addition, to obtain physical interpretations in the quantum
 theory of constrained dynamical systems such as the ones we
 are considering, one has to deparametrize the theory and express
 phase space variables (such as the 3-metric and extrinsic curvature) in terms
 of the Dirac operators and the ``time''variable. (see e.g.
 \cite{carlo,t:diss,atu:tII}).}.
In all the models considered here, the generator of the ``time translation''
defined by the conditional symmetry $t^A$ is one such observable: its classical
analog is simply $p_0-p_+$, the momentum corresponding to the conditional
symmetry. However, since these models have two (configuration) degrees of
freedom, the reduced phase space is four dimensional whence one expects there
to exist four independent Dirac observables. The open question therefore is
that of finding the three remaining observables. In the type I model, there is
no potential, whence every generator of the Lorentz group defined by the flat
supermetric $\eta^{AB}$ is a  Dirac observable. Thus, it is easy to find the
complete set both classically and quantum mechanically. In the type II model,
the task is already made difficult by the presence of a non-zero potential.
However, now the potential is rather simple and can actually be eliminated by a
suitable canonical transformation \cite{atu:tII}. (In full general relativity,
canonical transformations with the same property exist \cite{a:new,tate:poly}.
However, the resulting supermetric is curved. In the type II model, the new
supermetric is again flat; only the global structure of the constraint surface
is different from that in the type I model.) In terms of these new canonical
variables it is easy to find the full set of Dirac observables both classically
and quantum mechanically. Thus, in the type I and II models, the issue of the
physical interpretation is under control. In the type VI$_0$ and VII$_0$
models, on the other hand, the issue is wide open. Resolution of this issue
will also shed light on the question of uniqueness of the inner product and
clarify the issue of singularities discussed above.

What is the situation with respect to the more complicated Bianchi models,
types VIII and IX? Now, the conditional symmetry which played a key role in
quantum theory no longer exists: the origin of this symmetry can be traced to
the presence of a diagonal automorphism group which happens to be zero
dimensional in type VIII and IX models (see references in footnote 2). However,
it is possible that these models admit some ``hidden'' symmetries --i.e.
symmetries which are not induced by space-time diffeomorphisms. There is indeed
a striking feature along these lines, first pointed out by Charlie himself
\cite{m:mini}: in the type IX models, there exists an asymptotic constant of
motion precisely at the early chaotic stage. Can one use it for quantization?

\newpage
\begin{center}
{\bf Acknowledgements:}
\end{center}
The authors would like to thank Jorma Louko for discussions and for
reading the manuscript.

%\bibliography{cosm}

\bibliographystyle{plain}

\end{document}